\documentclass[aps,amssymb,amsfonts,floatfix,showpacs,superscriptaddress]{revtex4}

\usepackage[utf8]{inputenc}
\usepackage[french,english]{babel}
\usepackage{graphicx}  
\usepackage{dcolumn}   
\usepackage{amssymb}   
\usepackage{amsmath}
\usepackage{epsfig}
\usepackage[T1]{fontenc}
\usepackage{appendix}
\usepackage{color}
\usepackage{siunitx}
\usepackage{cleveref}
\usepackage[normalem]{ulem}
\usepackage{soul}
\usepackage[dvipsnames]{xcolor}
\usepackage{txfonts}
\DeclareMathAlphabet\mathbfcal{OMS}{cmsy}{b}{n}

\usepackage{lipsum}

\graphicspath{{./figures/}}
\Crefname{figure}{Figure}{Figures}
\crefname{figure}{Fig.}{Figs.}
\Crefname{equation}{Equation}{Equations}
\crefname{equation}{Eq.}{Eqs.}
\crefname{table}{Table}{Tables}

\DeclareSIUnit\pxl{pixels}
\DeclareSIUnit\ppm{ppm}
\DeclareSIUnit\ppb{ppb}
\DeclareSIUnit\ppt{ppt}
\DeclareSIUnit\mu{\micro}
\DeclareSIUnit{\liter}{$\ell$}

\begin{document}

\title{\emph{Ab-initio} calculations of laser-atom interactions reveal harmonics feedback during macroscopic propagation}

\author{{N. Berti}}
\affiliation{GAP, Universit\'e de Gen\`eve, Chemin de Pinchat 22, 1211 Geneva 4, Switzerland}
\author{{P. B\'ejot}}
\affiliation{Laboratoire Interdisciplinaire CARNOT de Bourgogne, UMR 6303 CNRS-Universit\'e de Bourgogne, BP 47870, 21078 Dijon, France}
\author{{Eric Cormier}}
\affiliation{Centre Lasers Intenses et Applications, Universit\'e de Bordeaux-CNRS-CEA, UMR 5107, 351 Cours de la Libération, F-33405 Talence, France}
\author{{J. Kasparian}}
\email{jerome.kasparian@unige.ch}
\affiliation{GAP, Universit\'e de Gen\`eve, Chemin de Pinchat 22, 1211 Geneva 4, Switzerland}
\affiliation{Institute for Environmental Sciences, University of Geneva, bd Carl Vogt  66, 1211 Geneva 4}
\author{{O. Faucher}}
\affiliation{Laboratoire Interdisciplinaire CARNOT de Bourgogne, UMR 6303 CNRS-Universit\'e de Bourgogne, BP 47870, 21078 Dijon, France}
\author{{J.-P. Wolf}}
\affiliation{GAP, Universit\'e de Gen\`eve, Chemin de Pinchat 22, 1211 Geneva 4, Switzerland}

\date{\today}

\begin{abstract}
We couple the full 3D \emph{ab initio} quantum evolution of the light pulse polarization in interaction with an atom with a propagation model to simulate the propagation of ultrashort laser pulses over macroscopic dimensions, in the presence of self-generated harmonics up to order 11.
We evidence a clear feedback of the generated harmonics on propagation, with an influence on the ionization probability as well as the yield of the harmonic generation itself.
\end{abstract}


\maketitle

\section{Introduction}
The propagation of high-intensity, ultrashort laser pulses gives rise to a wide variety of nonlinear processes, from higher-harmonic generation (HHG)~\cite{Li1989,Brabec2000,Agostini2004,Krausz2009} to filamentation~\cite{ChinHLLTABKKS2005,berge2007ultrashort,couairon2007femtosecond,Wolf2018}. Most descriptions of the processes rely on the single atom picture as in the so-called 3-step model~\cite{corkum1993plasma,Schafer1993}. Propagation effects through the target are often ignored or modeled as a macroscopic phase matching between the fundamental and the harmonics~\cite{rundquist1998phase,Popmintchev2009}, neglecting phenomena like the interplay between the generated harmonics and the non-linear propagation of the high intensity driving pulse.

An exact treatment combining the time-dependent \emph{ab-initio} Schrödinger equation (TDSE) description of the interaction of the electric field with atoms and the Maxwell solving of the non-linear propagation is extremely complex. 
 For this reason, simplified  microscopic models like Strong-Field Approximation (SFA)~\cite{Lewenstein1994,Priori2000,Gaarde2008}, quantum re-scattering~\cite{Le2009,Jin2011}, or the metastable state theory~\cite{Kolesik2014,Bahl2017} were developed to make the entire numerical processing tractable. 

Filamentation and harmonic generation have long been treated independently although they occur at similar intensities and rely on nonlinear atom-field interactions. Still, efficient third harmonic (TH) generation with a conversion efficiency up to a few percent was measured in laser filaments~\cite{liu2011efficient,gaarde2009intensity,berge2005supercontinuum,akozbek2002third,vockerodt2012low,kolesik2006simulation}, even on distances as short as a few mm~\cite{rundquist1998phase}. Harmonics up to the 23$^\text{rd}$ order have been observed in filaments in argon~\cite{steingrube2011high,vockerodt2012low}, at intensities in the \SI{100}{TW/cm^2} range at \SI{800}{nm}. 

Recently, the TH and its relative phase were shown to significantly alter the ionization probability in laser filaments at \SI{800}{nm}~\cite{bejot2014harmonic,bejot2015subcycle,doussot2016impact}. The TH also affects the nonlinear refractive index~\cite{perelomov1967ionization}, possibly saturating the Kerr effect~\cite{bejot2011higher,bache2012higher} and affecting phase matching that is crucial to the HHG yield~\cite{Paul2003,Gibson2003,weerawarne2015higher}. Harmonics therefore seem to have a non-negligible feedback on the filamentation dynamics. Despite this potential impact, they are often omitted in propagation codes. To the best of our knowledge, no work to date has included harmonics beyond the $5^{th}$.

Contradictory results about the higher-order Kerr effect (HOKE) in air~\cite{loriot2009measurement,loriot2010measurement}, that raised both positive~\cite{kolesik2010femtosecond,bache2012higher,petrarca2012higher,richter2013role,weerawarne2015higher} and negative reactions~\cite{odhner2012ionization,wahlstrand2012absolute,kohler2013saturation,kosareva2011arrest}, showed the limitations of modeling of filamentation on the basis of a simple empirical parametrization of the medium polarization and ionization~\cite{mitrofanov_mid-infrared_2015}.
A comprehensive picture requires to describe the full quantum polarization response of the medium to the electric field~\cite{rensink2014model,kohler2013saturation,volkova2013nonlinear,schuh2016interaction,lu2001moving,bandrauk1993exponential,bandrauk2009quantum}, including a saturation or an inversion of the Kerr effect~\cite{nurhuda2002ionization,nurhuda2008generalization,kano2006numerical,kohler2013saturation} as well as a closure of ionization channels due to the Stark effect~\cite{bejot2013high,wiehle2003dynamics,cormier2001above}.
Similarly, the impact of various harmonics, including low-order ones, on the generation of the HHG themselves, has recently been evidenced~\cite{Krueger2016,Orenstein2017,Hofmann2018}.

So far, very few  \emph{ab initio} microscopic models are fully coupled with macroscopic propagation codes. In fact, global descriptions of ultrashort pulse propagation including harmonic generation is necessary to overcome the limitations of the above-mentioned simplified models. Such global descriptions have been proposed along the years~\cite{lorin2007numerical,lorin2011wasp,Lorin2012,schuh2014simple,lorin2015development}. However, the computing requirements limited them to \SI{10}{\micro m} propagation distance~\cite{lorin2011wasp}.
Here, we have implemented an ab-initio description of the atom and a semi-classical representation of the interaction, and coupled it to a propagation model over \SI{5}{mm}, a distance in line with standard gas jet interaction lengths in HHG experiments~\cite{Li1989,Brabec2000,Agostini2004,Krausz2009}. By filtering out selected spectral ranges, we showed that harmonics up to the 9$^\text{th}$ impact on both the ionization process and the lower-order harmonics generation and therefore have a significant feedback on the pulse propagation.

\section{Methods}

We performed one-dimensional, time-resolved simulations (1D+1) of the propagation of a laser pulse initially centered at \SI{800}{nm}, with a peak intensity of \SI{50}{TW/cm^2} and a full width at half maximum (FWHM) of \SI{20}{fs}, in atomic  hydrogen at atmospheric pressure. 

To include the quantum calculation of the polarization in the propagation equation, we start with a real-field propagation equation
	\begin{equation}
		\label{eq_prop1_bis}
		\Delta \mathbf{E} - \dfrac{1}{c^2} \partial_t^2 \mathbf{E} = \mu_0 ( \partial_t \mathbf{J} + \partial_t^2 \mathbf{P} ).
	\end{equation}
where $\mathbf{E}$ is the electric field, $c$ is the speed of light, the polarization $\mathbf{P}$ is obtained from the bound states of the atomic wavefunction, and the current density $\mathbf{J}$ from its \textit{continuum} states. We then define $\mathbfcal{P}$ to gather them in a single term in Eq.~(\ref{eq_prop1_bis}):
	\begin{equation}
		\Delta \mathbf{E} - \dfrac{1}{c^2} \partial_{t}^2 \mathbf{E} = \mu_{0} \partial_{t}^2 \mathbf{\mathbfcal{P}}  .
		\label{eq:9}
	\end{equation}		
Considering a field linearly polarized along $\mathbf{z}$, propagating along $\mathbf{x}$, we can express this equation in the momentum space as :
\begin{equation}
	k_x^2 \widetilde{E}_z - \dfrac{\omega^2}{c^2} \widetilde{E}_z = \mu_0 \omega^2 \, \widetilde{\mathcal{P}}_z \ .
\end{equation}
where $k_x$ is the wave vector and $\omega$ is the frequency. Then, the field can be described as :
\begin{equation}
	\widetilde{E}_z = \dfrac{\mu_0 \omega^2}{k_x^2 - \frac{\omega^2}{c^2}} \, \widetilde{\mathcal{P}}_z \ .
\end{equation}
Considering only the forward-propagating field, we can keep only the right term of:
\begin{equation}
\dfrac{1}{k_x^2 - \frac{\omega^2}{c^2}} = \dfrac{1}{2 \frac{\omega}{c}} \Bigg( \dfrac{1}{k_x - \frac{\omega}{c}} - \dfrac{1}{k_x + \frac{\omega}{c}} \Bigg) .
\end{equation}
In this condition, the propagation equation become: 
\begin{equation}
	-k_x \widetilde{E}_z = \dfrac{\omega}{c} \, \widetilde{E}_z + \dfrac{\omega}{2 \epsilon_0 c} \, \widetilde{\mathcal{P}}_z \ ,
\end{equation}
 $\epsilon_{0}$ being the permittivity of vacuum. We define the vector potential $\mathbf{A}$ such that $\mathbf{E}=-\partial_t \mathbf{A}$. In the Fourier space, $\widetilde{E}_z = \textrm{i} \omega \widetilde{A}_z$, so that: 
\begin{equation}
	- \textrm{i} k_x \widetilde{A}_z = \partial_x \widetilde{A}_z = \textrm{i} \dfrac{\omega}{c} \, \widetilde{A}_z + \dfrac{1}{2 \epsilon_0 c} \, \widetilde{\mathcal{P}}_z
\end{equation} 
Moving to the referential moving at the pulse velocity:
\begin{equation}
	\partial_x \widetilde{A}_z = \textrm{i} \, \Big( \dfrac{\omega}{c} - \dfrac{\omega}{v} \Big) \, \widetilde{A}_z + \dfrac{1}{2 \epsilon_0 c} \,  \widetilde{\mathcal{P}}_z \ .
\end{equation}
If we define the polarization as $\widetilde{\mathcal{P}}_z = \widetilde{\varmathbb{P}}_z + \textrm{i} \omega \epsilon_0 \rho_{\textrm{at}} \alpha_0 \widetilde{A}_z$, ($\rho_{\textrm{at}}$ being the density of atoms in the propagation medium and $\alpha \simeq 4.593 \, \text{a.u.}$ the hydrogen polarizability at 800~nm) we can write the previous equation as:
\begin{equation}
	\partial_x \widetilde{A}_z = \textrm{i} \Big(\dfrac{\omega}{c} - \dfrac{\omega}{v} \Big) \, \widetilde{A}_z  + \dfrac{\widetilde{\varmathbb{P}_z} + \textrm{i} \omega \epsilon_0 \rho_{\textrm{at}} \alpha_0 \widetilde{A}_z} {2 \epsilon_0 c} , 
\end{equation}
then
\begin{equation}
	\partial_x \widetilde{A}_z = \textrm{i} \, \Bigg[ \dfrac{\omega}{c} \Big(1 + \dfrac{\rho_{\textrm{at}} \alpha_0}{2}  \Big)  - \dfrac{\omega}{v}  \Bigg] \, \widetilde{A}_z + \dfrac{1}{2 \epsilon_0 c} \, \widetilde{\varmathbb{P}}_z.
\end{equation}
By definition $\rho_{\textrm{at}} \alpha_0 = \chi_0^{(1)}$ and $1+ \dfrac{\chi_0^{(1)}}{2} \approx \sqrt{1 + \chi_0^{(1)}} = n_0$, if $\chi_0^{(1)}~\ll~1$. Therefore, the propagation equation rewrites:
\begin{equation}
	\partial_x \widetilde{A}_z = \textrm{i} \, \Big( \frac{n_0}{c} \omega - \dfrac{\omega}{v}  \Big) \, \widetilde{A}_z + \dfrac{1}{2\epsilon_0 c} \, \widetilde{\varmathbb{P}}_z.
\end{equation}

Since the electric field is linearly polarized along $\mathbf{z}$, the propagation equation in the referential moving at the phase velocity $v_{\text{p}} = c / (1+ \frac{\rho_{\text{at}} \alpha }{2})$ reduces to the unidirectional pulse propagation equation (UPPE) \cite{kolesik2002unidirectional,berti2015nonlinear}:
\begin{equation}
\partial_x \widetilde{A}_z(\omega) = \dfrac{1}{2 \epsilon_0 c}(\widetilde{\mathcal{P}_z}(\omega) - \textrm{i} \, \omega \epsilon_0 \rho_{\textrm{at}} \alpha \widetilde{A}_z(\omega)),
\label{UPPE}
\end{equation}

\begin{figure}[t]
		\centerline{
			\includegraphics[width=0.6\columnwidth]{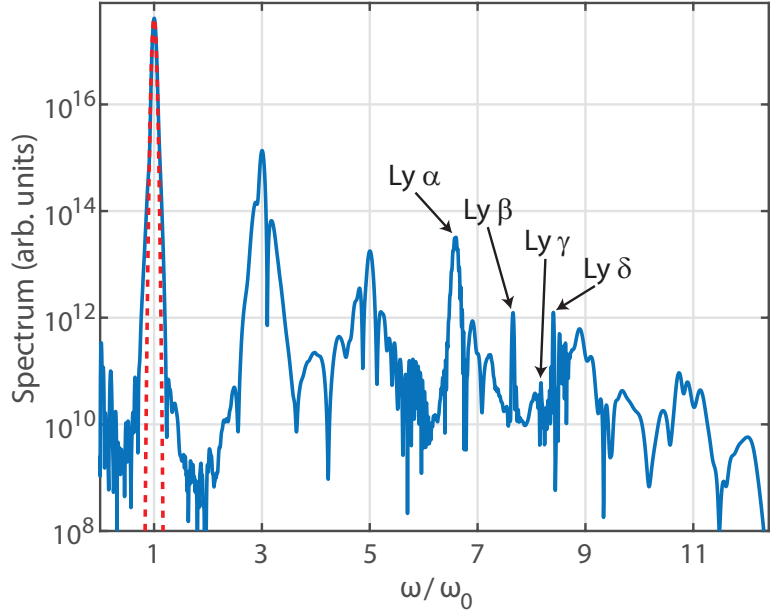}	
				}	
		\caption{Electric field spectrum of a \SI{50}{TW/cm^2}, \SI{20}{fs} (FWHM), Gaussian pulse after propagation over \SI{5}{mm} in hydrogen. Ly denote the Lyman lines. The dotted red line displays the initial spectrum.}
		\label{Fig0}
\end{figure}	

The polarization  $\mathbfcal{P}$ is deduced from the wavefunction $\Psi (\mathbf{r},t)$ of the atom under the influence of an electric field $\mathbf{E}(t)$, linearly polarized  along $\mathbf{z}$. We use the 3D TDSE in the velocity gauge~\cite{tannoudji1973mecanique}:
	\begin{equation}
		\text{i} \hbar \frac{\partial}{\partial t} \Psi(\mathbf{r},t) = \Big[  \frac{\bf{p}^2}{2m_e} + V(r) - \frac{q}{m_e} \mathbf{A}(t) \cdot \text{\bf{p}}  \Big] \Psi(\mathbf{r},t) ,
	\end{equation}
where $\text{\bf{p}}$ is the momentum. In atomic units, this equation displays as:
	\begin{equation}
		\text{i} \frac{\partial}{\partial t} \Psi(\mathbf{r},t) =  [\text{{H}}_l + \text{{D}}(t)] \Psi(\mathbf{r},t) ,
		\label{TDSE_Eq}
	\end{equation}
where $\text{{H}}_l = {\text{\bf{p}}^2}/{2} + V(r)$ is the free Hamiltonian of the system, and $\text{D}(t) = \mathbf{A}(t) \cdot \text{\bf{p}}$.  The angular dependence of the wavefunction is expressed in terms of spherical harmonics, while its radial part is discretized on a B-spline basis:
	\begin{equation}
		\Psi(\mathbf{r},t) = \sum_{l=0}^{l_\text{max}} \sum_{i=0}^{N} c_i^l(t) \frac{B_i^k(r)}{r} Y_l^0(\theta,\phi) .
	\end{equation}
Injecting this form into \cref{TDSE_Eq}, we get the matrix form of the coupled equation system:
	\begin{equation}
		\text{i} \text{\bf{S}} \cdot \frac{\partial}{\partial t} \text{\bf{c}}(t) = (\text{\bf{H}}_l +  \text{\bf{D}}(t)) \cdot \text{\bf{c}}(t) ,
		\label{TDSE_Eq_c}
	\end{equation}
	where the matrix $\text{\bf{S}}$ stems from the non-orthogonality of the B-spline functions:
	\begin{equation}
		s_{i,j} = \langle B_j^k |B_i^k \rangle 
				= \int_0^{\infty} B_i(r)B_j(r)dr .
	\end{equation}
and $\text{\bf{c}}(t)$ is the projection of $\Psi$ on $Nl_{\textrm{max}}$ dimensions, that will be propagated thanks to a Crank-Nicholson scheme. This scheme involves the matrix $\text{\bf{M}}(t) = \text{\bf{S}} + \text{i} \left( \text{\bf{H}}_0 + \text{\bf{D}}(t)\right){\delta t}/{2}$, that is highly sparse and bloc-tridiagonal, so that the bi-conjugate gradient method with preconditioning~\cite{sleijpen1993bicgstab,sleijpen1994bicgstab} is highly efficient on a single processor.

We then solve the full TDSE at each propagation distance and temporal point, using a highly efficient $B$-spline basis \cite{bachau2001applications,cormier1997above}. In this basis the interaction Hamiltonian matrix is several orders of magnitude sparser as compared to the eigen-basis. Together with the biconjugate stabilized gradient (BiCGStab) algorithm~\cite{sleijpen1993bicgstab}, this reduces the calculation of polarization to a few hours on a single processor, allowing  coupling with propagation. \cref{UPPE} was integrated by using the split-step Fourier method \cite{weideman1986split}, with a $4^{\textrm{th}}$-order Runge-Kutta scheme. The spatial propagation step is $\textrm{d}x=\SI{15}{\micro\m}$. We used 2048 grid points per optical cycle at the fundamental wavelength (\SI{800}{nm}), corresponding to a time step of $\simeq$ \SI{1}{as} and a box size of \SI{450}{fs}. To keep adequate sampling of the wave carrier, we limited the frequency range to $12\ \omega_0$.

The knowledge of the atomic wavefunction $\Psi$ then allows to calculate the polarization $\mathcal{P}_z$ as  $q_\textrm{e} \, \rho_{\textrm{at}} \langle\Psi|z|\Psi\rangle  $, with $q_\textrm{e}$ the electron charge. Note that this polarization includes both free and bound electrons, as their respective contributions are calculated in the same term, without introducing the usual arbitrary distinction between the current of free charges $\mathbf{J}$ and polarization $\mathbf{P}$ (See \cref{eq_prop1_bis,eq:9}). Unlike phenomenological models, it also intrinsically includes quantum effects like the Stark effect, that strongly affect the ionization via channel closure~\cite{bejot2013high}.

\section{Results}

\begin{figure}[t]
			\includegraphics[width=0.7\columnwidth]{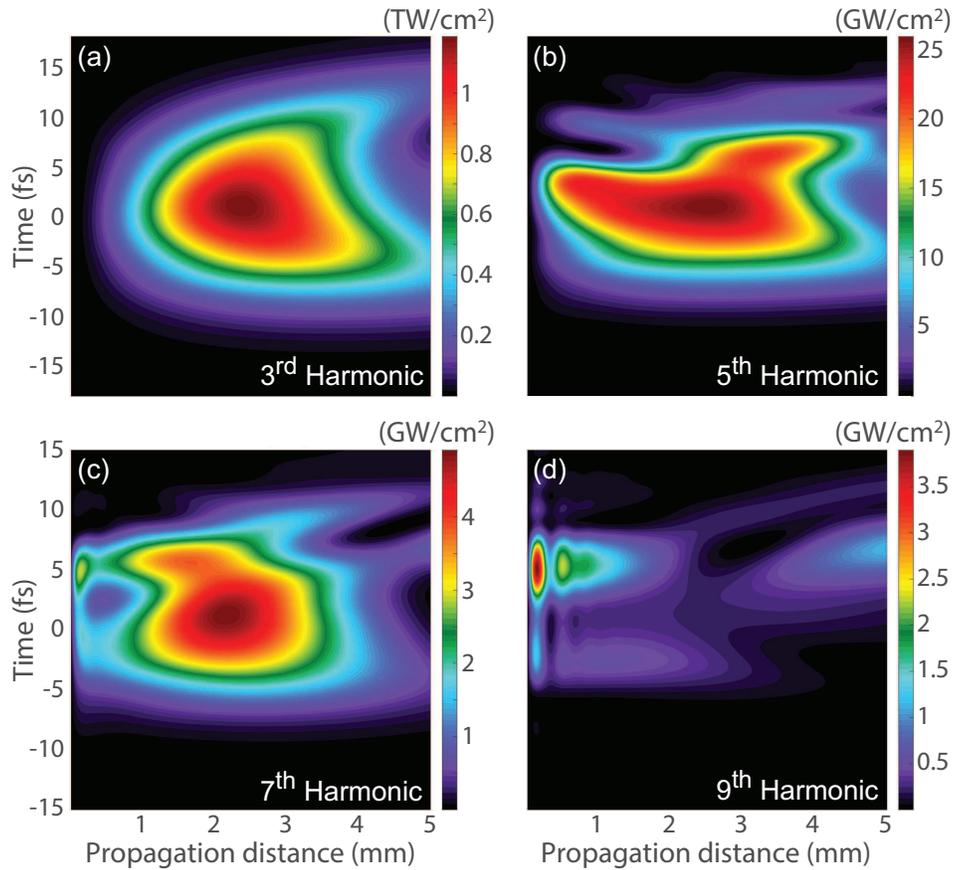}
		\caption{Temporal evolution of the harmonics along the propagation.}
		\label{Evolution_harmoniques}
\end{figure}

\begin{figure}[t!]
		\centerline{
			\includegraphics[width=0.7\columnwidth]{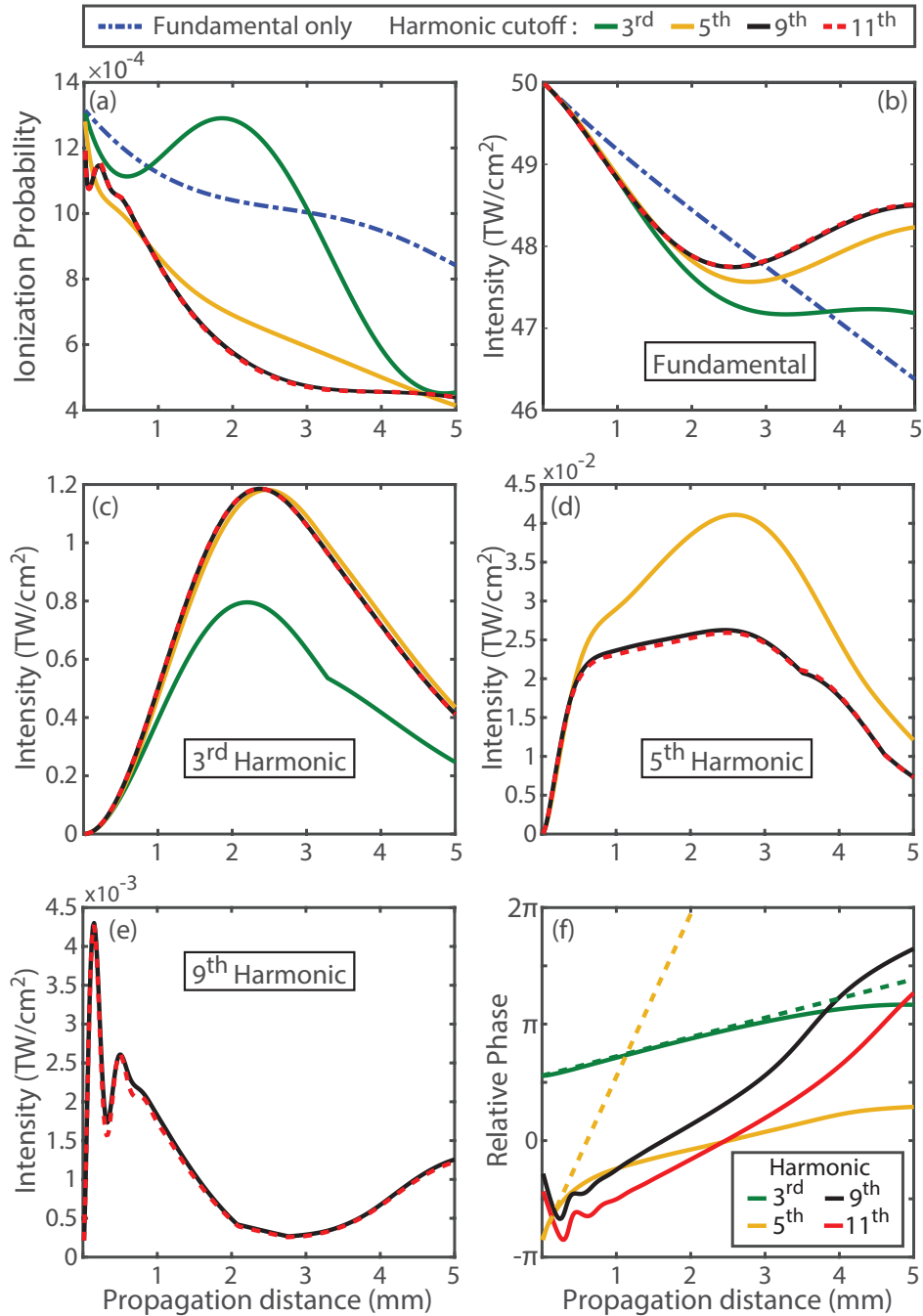}	
				}	
		\caption{Simulated propagation of a 50 TW/cm2, 20 fs pulse in atomic hydrogen. Ionization yield (a), intensity of the fundamental (b), 3rd (c), 5th (d), and 9th harmonic (e). (f) Relative phase of the harmonics with regard to the fundamental, calculated with the full spectral range. Dotted lines display the linear dispersion of the $3^\textrm{rd}$ and $5^\textrm{th}$ harmonics \cite{karplus1963variation}.}
		\label{Fig1}
\end{figure}

\Cref{Fig0} shows the electric field spectrum after propagating over \SI{5}{mm}. As expected at \SI{50}{TW/cm^2}~\cite{bandrauk2009quantum}, the harmonic yield decays by 1 to 2 orders of magnitude between two consecutive odd orders: 2\%, 0.08\%, and 0.008\% for the 3$^\textrm{rd}$, 5$^\textrm{th}$, and 9$^\textrm{th}$ harmonics, respectively, as also evidenced by the ranges of colorbars in \cref{Evolution_harmoniques,Fig1}(c)-(e).
The propagated pulse spectrum also includes the well-known Lyman emission lines between $6\ \omega_0$ and $9\ \omega_0$, especially where the $7^\textrm{th}$ harmonic is expected. 

These Lyman lines correspond to transitions from excited sates with levels $n > 1$ to the ground state $n=1$ (See \cref{Tab_Lyman}). At intensities of several tens of TW/cm$^2$, they are much more intense than the harmonics, and only one order of magnitude below the fundamental. 
Since the decoherence between excited and fundamental states is not considered in our model, the system continues to oscillate even after the field is switched off. 
As they travel slower than the fundamental frequency, the Lyman lines would be transferred from the positive (pulse trail) end to the negative (pulse front) end of the temporal box during the split-step calculation, be caught up by the fundamental pulse, and resonantly amplified. To prevent this, the Lyman frequencies are absorbed on the edge of the temporal numerical box with a soft filter (\emph{erf} function). This artificial damping forbids quantitative analyses of the Lyman lines evolution, and of their interaction with the remaining of the pulse, especially the $7^{th}$ and $9^{th}$ harmonics.

	\begin{table}[tb]
	\begin{center}
		\caption{\label{Tab_Lyman} The Lyman series}
		\begin{tabular}{|l|c|c|c|} 
		\hline
Notation   &   Transition   &   $\lambda$ (nm)  &  Order ($\omega/\omega_0$) \\		
		\hline
$\alpha$-Lyman   &  $n=2$ to $n=1$  & 121.5 	& 6.58 \\
\hline
$\beta$-Lyman   &  $n=3$ to $n=1$  & 102.5 	& 7.80  \\
\hline
$\gamma$-Lyman   &  $n=4$ to $n=1$   & 97.2 	& 8.23 \\
\hline
Limit   &  $n\to\infty$ to $n=1$  &  91.15	&  8.78 \\
\hline

		\end{tabular}
		\end{center}
	\end{table}

\Cref{Evolution_harmoniques} displays the evolution of the temporal shape of the harmonics along the propagation. The fundamental intensity decays linearly till \SI{2.5}{mm} and then re-grows (\Cref{Fig1}b). Conversely, harmonics 3, 5, and 7 first grow over approximately \SI{2.5}{mm}, and then decay. In contrast, harmonic 9 reaches a peak after as little as \SI{0.2}{mm}, then decays almost fully till \SI{3}{mm}, before entering in a new growth cycle. The faster rise and decay of the $7^{th}$, and even more of the $9^{th}$ harmonics, can be expected to stem from their partial overlap with the Lymann resonances, that strongly affect both their absorption, and, via Kramers-Kronig relationships, their dispersion.  

The evolution of the harmonic temporal pulse shape is more marked for the higher-order harmonics.  The $3^\textrm{rd}$ harmonic remains rather bell-shaped over more than \SI{3}{mm}, before a temporal shift accompanied with a slight pulse splitting. The $5^\textrm{th}$ harmonic self-shortens downto as little as less than \SI{10}{fs}, before splitting during the first mm of propagation and around \SI{4}{mm}. Pulse splitting is even more obvious in the $7^\textrm{th}$ harmonic, both around \SI{1}{mm} and from \SI{4}{mm} on, with even three sub-pulses at the end of the simulation. The $9^\textrm{th}$ harmonic  splits from the very first steps of the propagation, and the two sub-pulses merge back after \SI{2.5}{mm}. We attribute the complex evolution of the harmonics pulse shape to their wide band associated with their short duration. As a consequence, group-velocity dispersion is significant within each harmonic. The phase matching and the associated (back-) conversion efficiency are therefore different among spectral slices within each harmonic. 

To get more insight into the impact of each harmonics on the propagation and  ionization, we applied a low-pass filter with different cut-off frequencies to the polarization, preventing the generation and propagation of wavelengths beyond $2\omega_0$, $4\omega_0$, $6\omega_0$, $10\omega_0$, $12\omega_0$, i.e., respectively allowing propagation of the fundamental only, and of harmonics up to the $3^{\text{rd}}$, $5^{\text{th}}$, $9^{\text{th}}$, and $11^{\text{th}}$. Note that the effect of the $7^\textrm{th}$ harmonics on the propagation has not been investigated due to potential interferences, which cannot be adequately quantified, with the overlapping Lyman transitions lines (See \cref{Fig0}).

The ionization probability strongly depends on the harmonics taken into account in the propagation (\cref{Fig1}(a)). For example, considering harmonics up to the $11^{\text{th}}$ reduces the ionization probability by a factor 2, as compared with simulations considering the fundamental frequency only. 
The harmonics intensity is insufficient to yield any significant ionization by themselves. They contribute to the ionization via quantum interferences between pathways implying only photons at the fundamental wavelength, and pathways implying, e.g., one harmonic photon, of higher energy, beside fundamental photons, as was evidenced experimentally in the case of the third harmonic~\cite{doussot2015phase}. This effect therefore depends on both the intensity of the harmonics 
(\cref{Fig1}(c)-(e)) and their phases relative to each other and to the fundamental (\cref{Fig1}(f)), as discussed below in more details. Note that these phases do not depend much on the different spectral truncations. The TH contribution explains the oscillation of the ionization probability over the propagation, as the $3^{\text{rd}}$ harmonic is generated up to \SI{0.8}{W/cm^2} and subsequently decays (\cref{Fig1}(c)).

The fast oscillation of the ionization yield over the first \SI{600}{\micro\m} when considering the $9^{\text{th}}$ and $11^{\text{th}}$ harmonics also clearly correlates with that of the intensity of the $9^{\text{th}}$ harmonic, as well as the relative phase of the $9^{\text{th}}$ and $11^{\text{th}}$ harmonic during the propagation (\cref{Fig0}(f)).
All in all, depending on the range of harmonics considered in the calculation, the  ionization yield can vary by a factor of more than 2 after \SI{2}{mm} propagation, clearly evidencing that the harmonics are not a byproduct of the propagation, but rather actively contribute to its dynamics, at least up to the $9^\text{th}$. In contrast, the $11^\text{th}$ harmonics with only \SI{0.002}{\%} conversion efficiency has no influence on the ionization yield, although a  single 11$^\textrm{th}$ harmonic photon has an energy of \SI{17}{eV}, sufficient to photoionize the hydrogen atom.

The presence of harmonics also affects the evolution of the fundamental (\cref{Fig1}(b)). Without
harmonics, the linear decay of the fundamental intensity is due to dispersion, as well as to losses due to the energy spent to excite and/or ionize the atoms. Accounting for harmonics reduces the ionization yield, but as expected it accelerates the intensity decay of the fundamental along the propagation, due to its conversion into harmonics. 
 
 \begin{figure}[t]
		\centerline{
			\includegraphics[width=0.7\columnwidth]{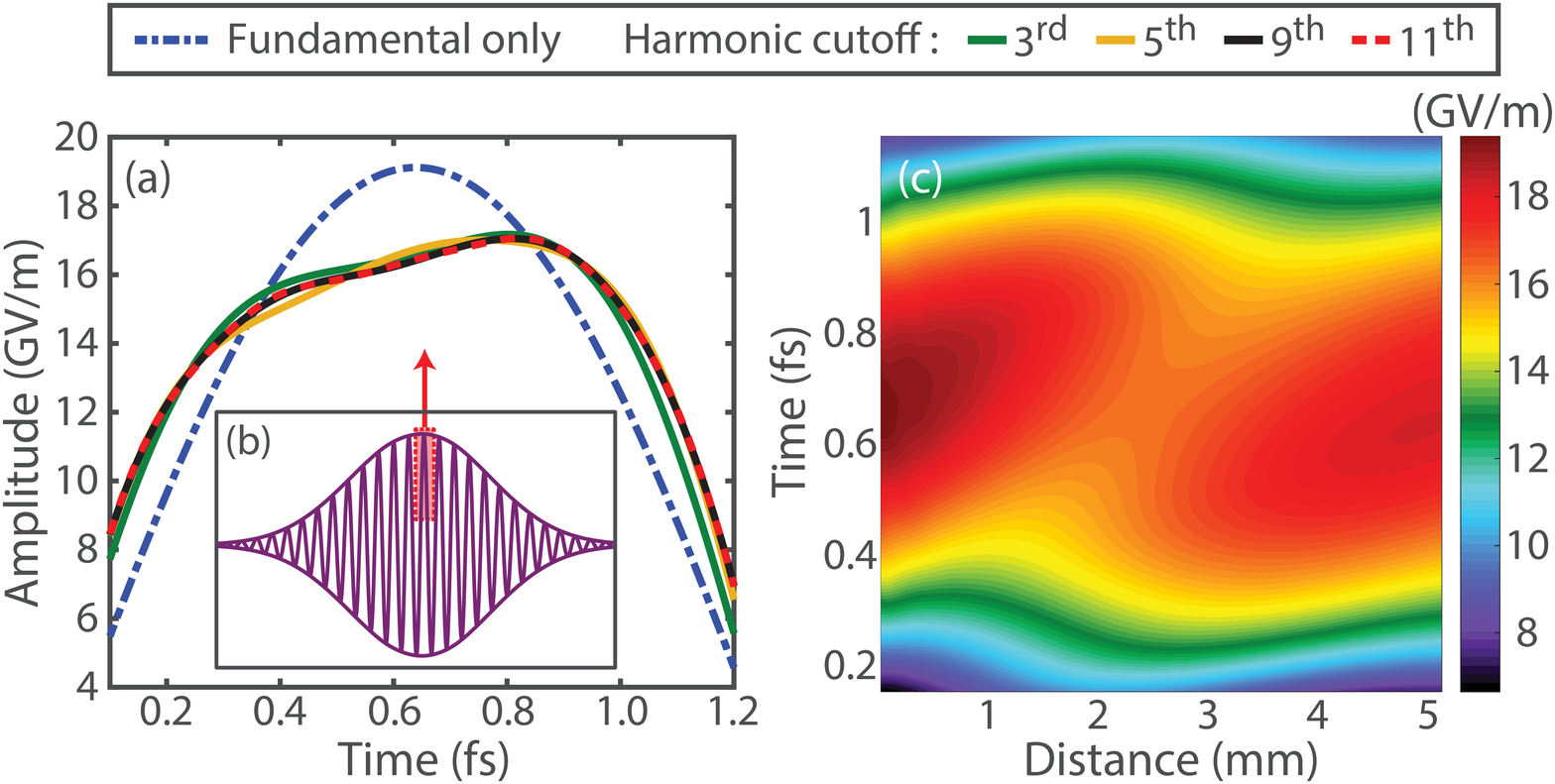}	
				}	
		\caption{(a) Electric field close to the centermost positive oscillation after $x=\SI{2}{mm}$ propagation, for simulations considering different spectral limit. (b) Overall pulse. (c) Evolution along the propagation of the centermost positive oscillation of the electric field.}
		\label{Fig4}
\end{figure}
 
Harmonics up to the 9$^{\textrm{th}}$ therefore substantially impact the propagation and deform the carrier wave, despite the very low conversion efficiencies. As shown in \cref{Fig4}(a), the carrier wave deformation is mostly explained by the contribution of the third harmonic. This is confirmed by \cref{Fig4}(c), where the envelope deformation is strong at the propagation distance (\SI{2}{mm}) where the TH intensity is close to its maximum. In spite of a minimal impact on the carrier wave, harmonics from the 5$^\textrm{th}$ have a substantial influence on ionization.

The feedback of the harmonics on the pulse propagation is also clearly visible on the harmonic generation itself, as well as on their temporal pulse shape (\cref{Fig1}(c)-(e)). Considering a broader spectral range covering the $5^{\text{th}}$ harmonic increases the yield of the $3^{\text{rd}}$ harmonic (\cref{Fig1}(c)). In contrast, considering the $9^{\text{th}}$ harmonics boosts the $3^{rd}$ harmonics while reducing the $5^{\text{th}}$ one (\cref{Fig1}(d)), that saturates already after \SI{0.5}{mm} of propagation and even decay beyond \SI{3}{mm}. Finally, the $11^{\text{th}}$ harmonic has little influence, if any, on the generation of the harmonics (\cref{Fig1}(c)-(e)), as was the case for ionization. This suggests that, in the considered conditions, the spectral range of the simulation does not need to be extended beyond the $9^\text{th}$ harmonics.

The ionization yield also depends on the relative phases $\Delta \phi_{\omega_n}= \phi_{\omega_n} - n\phi_{\omega_0}$ between the $n^{\textrm{th}}$ harmonics and the fundamental field, which modify the interferences between the above-mentioned ionization pathways. 
This can also be seen in terms of the influence of the harmonics on the carrier wave shape. In-phase (resp. out-of-phase) harmonics lead to a triangular (resp. square) optical cycles as expected from the Fourier series decomposition of triangle- and square-like periodic signals. Figure \ref{Fig1}(f) displays $\Delta \phi_{\omega_n}$ as a function of the propagation distance. At the beginning of the propagation, the third harmonic is weak (below 1\%) and in phase quadrature with the fundamental as expected from a three-photon process. As determined by Doussot \emph{et al.}~\cite{doussot2016impact}, this results in a lower ionization probability. However, when the third harmonic rises beyond 1\%, its contribution to ionization becomes positive, regardless of the phase~\cite{doussot2016impact}. But the monotonous increase of the phase during the propagation due to group-velocity dispersion finally induces back-conversion of the third harmonic, reducing its intensity, and, consequently resulting in a drop in the ionization yield.

The fifth harmonic is out-of-phase with respect to the fundamental field at the beginning of the propagation. This implies that the former is generated by the cascading process combining two photons of the fundamental field with a third harmonic photon: $5\omega_0=3\omega_0+\omega_0+\omega_0$.

The relative phases display wiggles that translate into fluctuations of the ionization probability (See \cref{Fig1}(f)). This complex evolution of the phases cannot be described by group-velocity dispersion only (See dotted lines for the 3$^\textrm{rd}$ and 5$^\textrm{th}$ harmonics) \cite{karplus1963variation}. It emphasises that harmonics follow a complex dispersion relationship influenced by the phase with which they are generated along the propagation, by the non-linear refractive index they encounter, as well as by the free electrons, in a process comparable to macroscopic quasi-phase matching relying on dispersion of a mixture of the neutrals and plasma~\cite{Paul2003,Gibson2003}.

Our simulations are consistent with experimental results displaying efficient harmonic generation in the propagation of high-intensity laser pulses~\cite{liu2011efficient,gaarde2009intensity,berge2005supercontinuum,akozbek2002third,vockerodt2012low,kolesik2006simulation,steingrube2011high}. Furthermore, the observed strong effect of the harmonics on ionization widely generalizes and extends previous experimental and numerical results on the third harmonic~\cite{bejot2014harmonic,doussot2016impact}. Note that the same approach could be used for, e.g., rare gases, by using a single-active electron potential instead of the purely Coulombian potential of the atomic hydrogen.

\section{Conclusion}

As a conclusion, a full coupled TDSE-UPPE propagation model has been developed and used to investigate the propagation of an ultrashort, high-intensity laser pulse over \SI{5}{mm}. Simulations over such distance has been made possible by an efficient \emph{ab-initio} TDSE solver based on $B$-spline basis set. 
The consideration of harmonics and their relative dephasing up to the $9^{th}$ strongly influences the propagation of ultrashort pulses, and especially their harmonic generation and ionization yield.
Our results illustrate that empirical models based on a complex envelope and phenomenological coefficients for ionization and the Kerr effect cannot adequately reproduce the complex behavior of the high-field -- atom interaction.
The emergence of single attosecond pulses may allow to probe the propagation at the sub-fs scale, ie, at the sub-optical-cycle scale~\cite{sommer2016attosecond}, therefore providing direct experimental comparison with our calculations, especially if the propagation is considered in two dimensions. Such $2D+z$ extension, implementing a transverse resolution, together with radial symmetry and efficient parallelization of the propagation steps, will allow its application to filamentation.

\acknowledgements{We warmly thank M. Moret and the Computer resource center (CRI) of the University of Burgundy for technical and numerical support, as well as J. Doussot for fruitful discussion. This work was supported by the ERC advanced grant "filatmo" and the Conseil Régional de Bourgogne (PARI program).}

\bibliography{MaBib}

\end{document}